\documentstyle[aps,prl,twocolumn,psfig]{revtex}

\newcommand{\pder}[2]{ \frac{\partial #1} {\partial #2} }

\newcommand{\com}[1]{ \: \mbox{#1} \quad }

\begin{document}

\draft
\preprint{not yet submitted}

\title{Quantum chaos with non-periodic, complex orbits
 in the Resonant Tunneling Diode }

\author{D.  S.  Saraga and T.  S.  Monteiro}
\address{ \,Dept.  of Physics and
Astronomy, University College, University of London, Gower St, London WC1E 6BT,
U.K.\\ } 
\date{\today}
\maketitle 

\begin{abstract}
We show that a special type of orbits, which are {\em non-periodic} and 
{\em complex}, 'Saddle Orbits' (SOs) describe accurately the quantal and 
experimental current oscillations in the Resonant Tunneling Diode in tilted 
fields. 
This is the first demonstration that one needs to abandon the 
periodic orbits (POs) paradigm in favour of this new and peculiar type of 
orbit. 
The SOs solve the puzzle of broad regions of experimental 
oscillations where we find no real or complex PO that can explain 
the data. 
The SOs succeeds in regimes involving several non-isolated POs, where 
PO formulas fail.
\end{abstract}


The Resonant Tunneling Diode (RTD) in tilted fields has recently been 
intensively investigated as an experimental probe of 'quantum chaos'
\cite{F94,M95,SS95,NS98a,W96,TD97}. 
It is widely considered to be a paradigm of Periodic Orbit (PO) Theory 
in a real system, 
yet to date no PO formula has been shown to provide a 
full and quantitative description of the current. 
Several approaches were presented recently \cite{N98,BR98,SM98}, 
expressing the tunneling current in terms of periodic orbits. 
We demonstrated previously \cite{SM99} that they could reproduce 
qualitatively the voltage range of experimental features 
like period doubling of the current. 
However, they only yielded reasonable agreement for the {\em amplitudes} 
of current oscillations in specific regimes, 
namely the stable (torus-quantization) region and its opposite extreme,
the isolated unstable periodic orbit regions. 
They failed in an intermediate regime spanning a broad range of field values. 
There one finds regions where there is no real PO 
or alternatively competing non-isolated POs. 
As we argue below, it is not simply a question of improving the PO theory 
with uniform approximations. 
In the regions where there is no real PO, even complex 'ghosts'\cite{KHD93} 
POs cannot explain the experimental oscillations 
(for convenience we continue nevertheless to refer to these regions as 
'ghost regions').

The main difficulty in establishing a semiclassical theory stems from the initial
state $\phi_0$ describing the electrons prior to tunneling: one cannot simply
exclude it from stationary phase considerations imposed on the rapidly varying
function $e^{i S(z,z')/\hbar}$ of the classical action $S$ of a trajectory. 
We show here that the correct stationary phase condition arising from the theory
proposed in \cite{BR98} yields orbits of a new type - which we call 'Saddle Orbits'
(SOs).
The SOs are {\em non-periodic} and {\em complex}, and describe the current
accurately even in regimes where the previous semiclassical PO theories failed.
They also show good agreement with the experimental amplitudes obtained from Bell
Lab data \cite{M95}. 
The SOs are quite distinct from the real closed orbits identified in atomic 
photoabsorption from localized ground states \cite{Delos}. 
As for ghosts, the imaginary component of the action provides a damping term.  
We show that for SOs a further weighting term due to $\phi_0$ can 
partially cancel this damping. 
This yields contributions which can decay slowly with decreasing $\hbar$, also
solving the puzzle of oscillations which persist far into the ghost regions.

We recall briefly the RTD model \cite{F94}. 
An electric field $F$ (along $x$) and a magnetic field $B$ in the $x-z$ plane (at 
tilt angle $\theta$ to the $x$ axis) are applied to a double barrier quantum well
of width $L=1200 \AA$. 
Electrons in a 2DEG
accumulate at the first barrier and tunnel through both barriers
giving rise to a tunneling current $I$. 
In the process they probe the classical dynamics -regular or chaotic- 
within the  well. 
The current oscillates as a function of applied voltage $V$.
After rescaling \cite{SM98} with respect to $B$, the dynamics at given $\theta$
and ratio of injection energy to voltage ($R=E/V \sim 0.15$ for the
Bell Lab experiments) depends only on the parameter
$\epsilon=V/LB^2$. 

The theoretical scaled current used for our quantal calculation and the starting
point of the semiclassical theory is expressed as a density of states weighted by a
tunneling matrix element:  $I(B) =\sum_i{W_i \delta(B-B_i)}$.
We used the Bardeen matrix element \cite{Bard61} form for $W_i$, 
which is an overlap between  $\phi_0$  and the wave function $\psi_i$ 
in the quantum well. 
In the experiments, incoherent processes
such as phonon emission damp the current by  $e^{-T/\tau}$, where $T$
is the period of the motion of an electron in the well and $\tau \sim 0.11$ ps.
Details of the experimental data reduction and quantum calculations were
given in \cite{SM98,SM99}. 
We can obtain reliable experimental amplitudes in regimes where 
the current is dominated by a single frequency (pure period-one or period-two). 

For the semiclassics, one can re-express the Bardeen matrix element 
in terms of energy Green's functions and use
their semiclassical expansion over classical paths to get the following expression
for the tunneling current \cite{BR98}:
\begin{equation}
\hspace{-3mm} I(B) \propto  \Re e  \int \! \! dz \! \int \! \! dz' \sum_{{\rm cl}}
 m_{12}^{-1/2} e^{i S(z,z')-B \cos{\theta} (z^2 +z'^2)/2}
\label{int}
\end{equation}
where $m_{12}=\frac{\partial z}{\partial p_{z_0}}$ is
an element of the monodromy matrix $m_{ij}$, for a trajectory 
$(x=0,z) \to (x'=0,z')$  which 
must connect both walls of the well.  
We consider here that the initial state is in the lowest Landau state: 
$\phi_0(z)=\sqrt{B \cos \theta /\pi} \exp(-B \cos \theta z^2/2)$. 
The stationary phase condition applied to Eq.(\ref{int}) gives 
$i \pder{S}{z} - B \cos\theta z = 0 =i \pder{S}{z'} - B \cos\theta z'$. 
The contributing trajectories $(z,p_z) \to (z',p_z')$ will therefore satisfy the
condition: 
\begin{equation} 
 p_z=i B \cos\theta z \com{,} p_z'= -i B \cos\theta z' 
\label{so}
\end{equation}
One sees clearly  that these trajectories, which we call 'Saddle Orbits' 
(SOs), will invariably be {\em complex} and {\em non-periodic}.
It follows that we cannot consider the repetitions of a given SO 
-as one does for POs when investigating period-doubled oscillations. 
Instead, one finds other SOs of quite different shapes with longer periods. 
We found that all the SOs which give a substantial contribution to the current have 
$z=z'$. 
They retrace themselves {\em once} because of an intermediate bounce normal 
to a wall or a 'soft' bounce on the energy surface. 

In \cite{BR98} the $B \cos\theta$ term in Eq. (\ref{so}) was neglected
in order to get a formula in terms of POs with null momentum $p_z=0=p_z'$.
One can establish a link between a self-retracing SO and its PO counterpart
by expressing the SO in term of an expansion around the PO.
Making a Taylor expansion of the action and
neglecting $\frac{\partial^n S} {\partial z^n} $ for $n>2$, one finds :

\begin{equation}
z_{SO}=\frac{ z_{PO} }{1-\delta}
\label{poapp}
\end{equation}
where $\delta= i \cos \theta \frac{ \bar{m}_{12} }{ \bar{m}_{11}-1 }$ 
and $\bar{m}_{ij}$ is the scaled monodromy matrix of the PO. 
We found that non self-retracing SOs, which correspond to segments of POs, 
are not relevant to these experiments.

\begin{figure}[htb]
\centerline{\psfig{figure=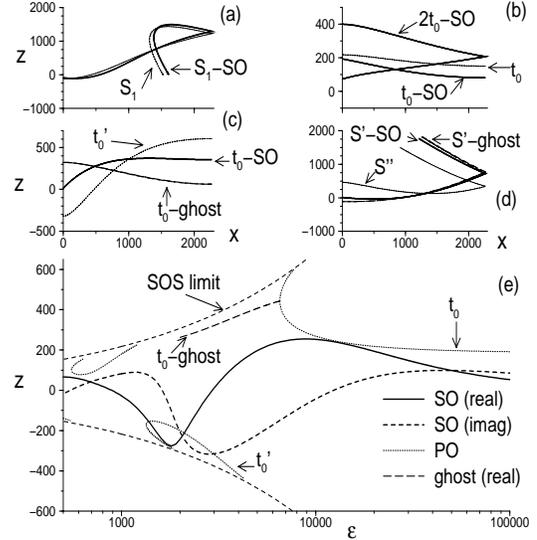,angle=270.,height=7.cm,width=7.cm}}
\vskip 0.5cm \caption{ a)-d) Shape in $x-z$ plane of the {\em real} part of SOs,
with the related POs. 
a) $\theta=27^\circ, \epsilon=2000$.
Differences between the SO and its counterpart PO ($S_1$) are minimal.
b) $\theta=11^\circ, \epsilon=20000$.
We show the main period-one PO $t_0$, its related SO as well as the
SO which is related to its the second repetition $2t_0$. 
c) $\theta=11^\circ, \epsilon=3000$. 
The SO is between the real PO $t_0'$ and the ghost PO $t_0$. 
d) $\theta=27^\circ, \epsilon=5000$.
A ghost PO ($S'$) and a real PO ($S''$) are present. 
The real part of the SO is closer in shape to the real part of the ghost.
e) $\theta=11^\circ$. 
Evolution of the starting $z$ (at $x=0$) with $\epsilon$ for the main
period-one PO $t_0$, showing part of its infinite cascade of 
tangent bifurcations where $t_0$ disappears, leaving a ghost while a
similar PO appears at a lower $\epsilon$.
Also shown is the behaviour of the related SO. 
We see that a single SO 'interpolates' smoothly between the successive 
disappearing and re-entrant POs. 
This illustrates the striking simplicity and power of the SO approach 
in comparison with POs.}
\label{pl-so}
\end{figure}

We plot some of these trajectories in Fig. \ref{pl-so}a)-d), together with their 
corresponding PO counterpart.
The link between SOs and POs is evident in Fig. \ref{pl-so} a),
which shows a regime where the SO and PO are very similar.
In this case the $x-z$ path of the SO is not very complex 
(the imaginary part is an order of magnitude smaller than the real part). 
The connection between SO and PO is generally not so apparent though. 
Fig. \ref{pl-so} b) shows the stable PO $t_0$, its related 'primitive' SO $t_0$-SO
as well as {\em another} SO, which seems completely unrelated to them.
This SO ($2t_0$-SO) plays, in fact, the role of the {\em second repetition} of 
the 'primitive' SO: its complex action and the real part of its period are twice 
(within $1\%$) those of $t_0$-SO. 
Despite the obvious difference in their path, some of the 
properties of the (SO-PO) pair such as their monodromy matrix 
or action can be very similar. 

Perhaps the most interesting situation is seen in Fig. \ref{pl-so} c), 
where the SO 'interpolates' between the ghost PO and the real PO.
This fact is illustrated in Fig. \ref{pl-so} e), where we plotted the evolution of
the starting position $z$ (at $x=0$) against $\epsilon$.
One sees that a single SO is linked to a large number of POs.
The latter are the basic traversing 2-bounce POs ($t_0,t_0',..$) to which 
have been attributed (in previous work) the period-one oscillations of the current. 
Their complicated dynamics has been well studied\cite{NS97,SM98}.
They undergo an infinite cascade of tangent bifurcations, where
they disappear leaving a ghost. 
Subsequently at some lower $\epsilon$ a similar PO reappears from the opposite 
edge of the Surface of Section.
Over some $\epsilon$ interval this re-entrant PO coexists with the ghost of the old PO.

In comparison, the corresponding  SO behaves in a much simpler way, 'interpolating'
between the successive POs.
For example for $\epsilon>10000$ the SO is related to $t_0$, which disappears 
in a tangent bifurcation at $\epsilon=6500$.
The SO remains close to the $t_0$  ghost down to $\epsilon=3000$, where it veers 
away towards the new orbit $t_0'$ which appeared from the edge at $\epsilon=4300$.
One sees this fact clearly in Fig. \ref{pl-so} c), where the SO is between the
 $t_0$ ghost and $t_0'$.
The same happens at $\theta=27^\circ$, where the 3-bounce PO $S'$ disappears in a 
tangent bifurcation at $\epsilon=7700$, leaving a ghost, while a similar PO 
$S''$ appears at $\epsilon=5500$. 
As seen in  Fig. \ref{pl-so} d), the SO is very close the ghost of $S'$, but it will
 soon approach the new PO at lower $\epsilon$.
Note that SOs never disappear in bifurcations as they are non-periodic, 
but rather when they 'miss a bounce' on the emitter wall because of the 
voltage drop.

The semiclassical current is given by the straightforward Gaussian integration 
resulting from the stationary phase approximation applied on (\ref{int}). 
After normalizing to the amplitude at $\theta=0^\circ$, the current due to one SO
 is approximated by a simple analytical formula:
\begin{eqnarray}
I(B) &=& \Re e \frac{ e^{B (i \tilde{S} -\cos\theta z^2) + i \mu \pi/2} }
{\sqrt{-\cos\theta \tilde{m}_{12} + \tilde{m}_{21}/\cos\theta  + 2 i \tilde{m}_{11}}} 
\label{semi}  
\end{eqnarray}
where $\mu$ is a Maslov index, $\tilde{S}$ and $\tilde{m}_{ij}$ are the 
scaled action and element of the classical monodromy matrix of the SO. 
One can also use the expansion in Eq.(\ref{poapp}) to approximate the SO current by
the related PO. 
Expanding the action $S$ of the SO around the PO up to second order,
one finds:
\begin{eqnarray}
I(B) & \simeq & \Re e 
\frac{ e^{B (i \bar{S} -\cos\theta \bar{z}^2 \frac{1}{1-\delta})+ i \bar{\mu} \pi/2} }
{\sqrt{-\cos\theta \bar{m}_{12} + \bar{m}_{21}/\cos\theta + 2 i \bar{m}_{11}}} 
\label{semiapp}  
\end{eqnarray}
where $\bar{S}$ and $\bar{z}$ are the scaled action and starting position of the PO.
This is exactly the PO formula presented in \cite{BR98} and tested in 
\cite{SM99}. 
So we see that the PO formula will give accurate results
provided that the higher derivatives of $S$ are small {\em and} 
that the expansion of the SO around the PO is justified. 
This latter point is the most relevant, as it is the prime reason 
why the PO formula fails in certain regions.

\begin{figure}[htb]
\centerline{\psfig{figure=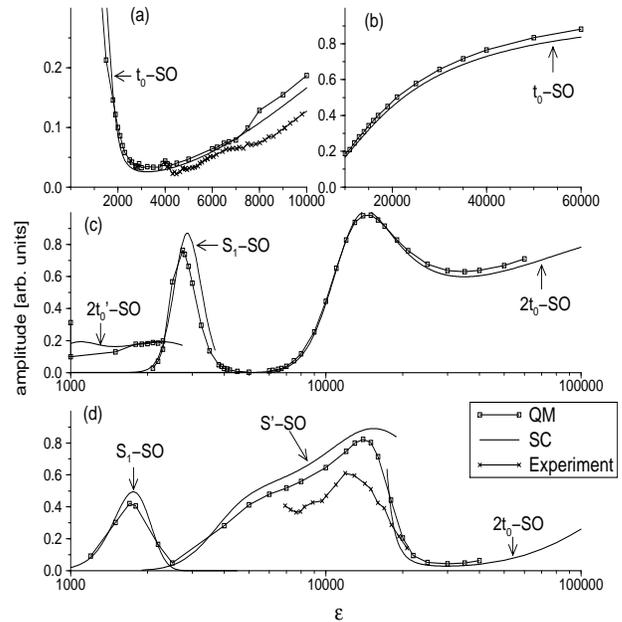,angle=270.,height=8.cm,width=8.cm}}
\vskip 0.5cm \caption {Quantal, experimental and semiclassical amplitudes. 
The SO labels indicate which PO they are related to.
Period one at $\theta=11^\circ$ in the unstable/ghost region (a) and 
in the stable (torus-quantization) regime (b). 
$t_0$-SO describes very accurately the quantal current, even in the ghost region
($2500 < \epsilon < 7000$) where PO theories fail.
(c) Period two at $\theta=11^\circ$. 
The SO formula improves over the PO formula, giving very 
accurately  the 
quantal maximum ($\epsilon \sim 15000$). Experimental amplitudes were not obtained
for this case since there was a strong period one beat in this region. 
(d) Period two at $\theta=27^\circ$.
As in (c), the peak at low $\epsilon$ is well described by both the
 SO and the PO formulae ($S_1$).
$S'$-SO gives the contribution to the very broad plateau where no real PO is present
($5000 < \epsilon < 8000$).
$2t_0$-SO is responsible for the current for $\epsilon > 17000$, with 
no overlap with $S'$-SO.}
\label{amp}
\end{figure}

We show in Fig. \ref{amp} a comparison between quantal, experimental and 
semiclassical amplitudes for period one and two currents. 
The corresponding comparison with the PO formula was presented in \cite{SM99}.

Fig. \ref{amp} a) shows the period one amplitudes at $\theta=11^\circ$ in the
unstable and 'ghost region', while Fig. \ref{amp} b) shows the stable torus regime.
We found \cite{SM99} that for the stable ($\epsilon>7000$) and the 
chaotic ($\epsilon<2500$) regions PO theory 
(using $t_0$ and $t_0'$) gave good agreement.  
These are regions where the SO and PO current are almost equal, 
according to Eq. (\ref{semiapp}). 
But the intermediate region ($2500 < \epsilon < 7000$) was a puzzle:
the PO formula including the ghost failed to account for the 
quantal and experimental oscillations, by a factor of $3$.
Fig. \ref{amp} a) shows that the SO completely solves this problem, giving
accurate results over the entire range from the torus regime to the chaotic 
regime, including the 'ghost region'.

The main reason why Eq.(\ref{semiapp}) fails in that region was illustrated in 
Fig. \ref{pl-so} a): the SO cannot be approximated by the ghost PO, as it is 
also related to the new $t_0'$. 
In this case it is not because the third derivative of $S$ is large.

The same happens for the $S'$ ghost at $27^\circ$ [Fig. \ref{amp} d)]. 
The SO describes very well the broad plateau of quantal and experimental 
amplitudes, while POs failed.
Once again, this is because the SO interpolates between the ghost 
(which appears at $\epsilon=7700$) and the new PO $S''$
(which appears at $\epsilon=5500$), as illustrated in Fig. \ref{pl-so} d).
Therefore, the failure of the PO theory in these regions is due to the sequence 
of tangent bifurcations, each followed by a new re-entrant PO,
[as illustrated in Fig. \ref{pl-so} f)], which rules out a simple connection 
between one SO and one PO.

The SO also solves the intricate superposition of two non-isolated 
(in action and phase-space localization) POs \cite{SM99}: 
the second repetition of $t_0$ (which yields quantized torus states) 
and $S'$ for $\epsilon \sim 13000$, $\theta=27^\circ$.
Indeed, there is no overlapping region for the related SOs, which now describe 
accurately the current, as shown in Fig. \ref{amp} d). 
In Fig. \ref{amp} c) one sees that the period doubling maximum around 
$\epsilon \sim 15000$ is described very precisely by the SO, an improvement
over the PO results. 
Finally, we note that for both angles the period two maxima at low  
$\epsilon$ $\sim 2000$ are described equally well by either the PO or 
the SO formula.
Here both trajectories ($S_1$-SO and $S_1$) are very similar as seen in
Fig. \ref{pl-so} c).
The experimental amplitudes are lower than theory, but this is consistent with 
a $10\%$ uncertainty in $\tau$.

The strength and persistence of the contribution of these complex orbits, 
even in the region where the SO formula does not reduce to the PO formula
 (such as in the ghost regions) is quite remarkable.
Usually (e.g., in the density of states or the photoabsorption 
spectra of atoms \cite{KHD93,D95}),
the contribution of complex ghost POs is extremely weak. 
They are exponentially damped 
away from the bifurcation (as $\epsilon$ changes),
since the imaginary part of the action increases as the 
ghost becomes more complex.
However, in the RTD the tunneling amplitude includes an additional term 
due to the initial state:
\begin{eqnarray}
|I(B)| & \propto &  e^{- B (\tilde{S_I} + \cos\theta (z_R^2-z_I^2))} 
\label{semamp}  
\end{eqnarray}
where the subscripts $I$ and $R$ denote the real and imaginary parts.
We see that the {\em imaginary} part $z_I$ of the starting position of a 
SO can compensate for the damping due to $\tilde{S_I}$.
This is the reason why even 'very' complex SOs can contribute.
For instance, the $S'$-SO amplitude in the ghost region 
at $\theta=27^\circ$ shows a broad plateau, and no exponential
damping away from the bifurcation ($\epsilon < 7700$).

\begin{figure}[htb]
\centerline{\psfig{figure=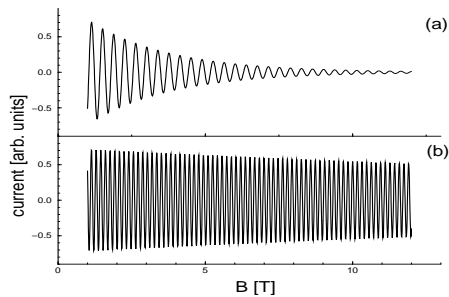,angle=270.,height=4.cm,width=6.cm}}
\vskip 0.5cm \caption {Semiclassical current $I(B)$ in two ghost regions
showing the unusual $\hbar$ dependence. 
(a) $\theta=11^\circ, \epsilon=4000$. 
The SO current is exponentially damped with $B$ (i.e., in the classical limit).
(b) $\theta=27^\circ, \epsilon=7000$. 
The current is only linearly damped with $B$, showing a persistence in the 
classical limit never seen in ghost POs.
}
\label{damp}
\end{figure}

Similarly, ghost POs are  exponentially damped in the 
classical limit $\hbar \to 0$
(which corresponds in our scaled model to $B \to \infty$).
In Fig. \ref{damp} we investigate the $\hbar \to 0$ behaviour
of the SO current in the two ghost regions.

In Fig. \ref{damp} a) ($\theta=11^\circ, \epsilon=4000$) 
the amplitude of $t_0$-SO is exponentially damped with $B$. 
This behaviour is seen in the experimental ghost region 
(see Fig. 5 of \cite{SM98}): along curves of constant $V/B^2$
the amplitudes decay rapidly with $B$. 
Fig. \ref{damp} b) ($\theta=27^\circ, \epsilon=7000$) shows 
a more surprising behaviour.
Here the amplitude of $S'$-SO is not exponentially but {\em linearly} damped. 
This very surprising feature is seen experimentally (see Fig. 6 of \cite{SM98}).
The explanation for the persistence of the complex SO in the classical limit
is found again in (\ref{semamp}).

We emphasize that this work is consistent with previous
studies of 'scarring' in the RTD.
It has been found that quantum states localized near some isolated or multiple POs
can dominate the tunneling \cite{W96,NS98a}.
We have also investigated wavefunctions and Wigner
distributions. 
We found that in the strong scarring regions, where
the relevant scarring PO is only  marginally unstable
\cite{NS98a} (e.g. for $\theta=27^\circ, \epsilon=10000$),
the real part of the SO is very close to the PO  -within the '$\hbar$' quantum 
uncertainty.
This can also be the case with scars carried by ghost POs 
(e.g. $\theta=27^\circ, \epsilon=7000$).
This is reasonable since after all it is the bundle of classical trajectories 
in the {\em neighbourhood} of POs and SOs which scars or carries the electrons.
Regions like the $\theta=11^\circ$ ghost region
where the SO and PO are really different in shape do not show strong scarring
by single states and the quantal current is carried by 
broad clusters of states.

We conclude that the SOs are a novel and  successful way
to approach semiclassical quantization of these types of chaotic systems.
We recall that SOs arise solely from the inclusion of
the initial state in the stationary phase condition.
Hence one could expect SOs to be potentially relevant in the description of the
expectation value of a quantal quantity (expressed as a  
density of states  weighted by some matrix element \cite{Eck})
if the observable in the matrix element is very localised and so 
varies as rapidly as $e^{iS/\hbar}$.

We are greatly indebted to E. Bogomolny for  
providing us with invaluable help with his semiclassical formalism 
and to G. Boebinger for providing his experimental data. 
T. S. M. acknowledges funding from the EPSRC. 
D. S. S. acknowledges financial support from the TMR programme.

\end{document}